\documentclass{PoS}

\usepackage{caption}
\title{Spectroscopy of the All-Charm Tetraquark}

\ShortTitle{Spectroscopy of the All-Charm Tetraquark}

\author{\speaker{V.~R.~Debastiani}\\
       Departamento de F\'{\i}sica Te\'orica and IFIC, Centro Mixto Universidad de Valencia-CSIC Institutos de Investigaci\'on de Paterna, Aptdo. 22085, 46071
Valencia, Spain\\
       E-mail: \email{vinicius.rodrigues@ific.uv.es}}

\author{F.~S.~Navarra\\
        Instituto de F\'{\i}sica,
Universidade de S\~{a}o Paulo, C.P. 66318, 05389-970 S\~{a}o Paulo, SP, Brazil\\
        E-mail: \email{navarra@if.usp.br}}

\abstract{
We use a non-relativistic model to study the mass spectroscopy of a tetraquark composed by $c \, \bar{c} \, c \, \bar{c}$ quarks in the
diquark-antidiquark picture.
By numerically solving the Schr\"{o}dinger equation with a Cornell-inspired potential, we separate the four-body problem into three two-body problems.
Spin-dependent terms (spin-spin, spin-orbit and tensor) are used to describe the splitting structure of the $c\bar{c}$ spectrum and are also extended to the interaction between diquarks. Recent experimental data on charmonium states are
 used to fix the parameters of the model and a satisfactory description of the spectrum is obtained. We find that the spin-dependent interaction is sizable in the diquark-antidiquark system, despite of the heavy diquark mass, and that the diquark has a finite size if treated in analogy to the $c\bar{c}$ systems.
We find that the lowest $S$-wave $T_{4c}$ tetraquarks might be below their thresholds of spontaneous dissociation into low-lying charmonium pairs, while orbital and radial excitations would be mostly above the corresponding charmonium pair threshold. These states could be investigated in the forthcoming experiments at LHCb and Belle II.}

\FullConference{XVII International Conference on Hadron Spectroscopy and Structure - Hadron2017\\
		25-29 September, 2017\\
		University of Salamanca, Salamanca, Spain}

\begin{document}

\section{Introduction}

The possible $cc\bar{c}\bar{c}$ states have gained attention recently due to the increasing fauna of exotics and the possibility of producing charmonium pairs in high energy facilities like LHC and KEK.
Due to the absence of light quarks, it is unlikely that these states are
meson molecules  and the tetraquark picture is favored.
Assuming a diquark-antidiquark structure, we study the $[cc][\bar{c}\bar{c}]$ tetraquark with a non-relativistic model \cite{Debastiani}.\\
\

\newlength\Colsep
\setlength\Colsep{10pt}

\noindent\begin{minipage}{\textwidth}
\begin{minipage}[c][6cm][c]{\dimexpr0.5\textwidth-0.5\Colsep\relax}
\indent
The $T_{4c}$ was first proposed by Iwasaki in 1975 \cite{Iwasaki}, just after the discovery of the $J/\psi$.
A few years later, Chao studied this tetraquark in the diquark picture including orbital momentum and spin-dependent forces \cite{Chao}. After that, only
quite few works have investigated the role of orbital momentum or radial excitations. We have thus chosen to focus on these features, extending a successful charmonium model \cite{model} in order to test the diquark assumption and the contribution of spin-dependent terms.
\end{minipage}\hfill
\begin{minipage}[c][6cm][c]{\dimexpr0.5\textwidth-0.5\Colsep\relax}
\centering
\includegraphics[width=0.75\textwidth]{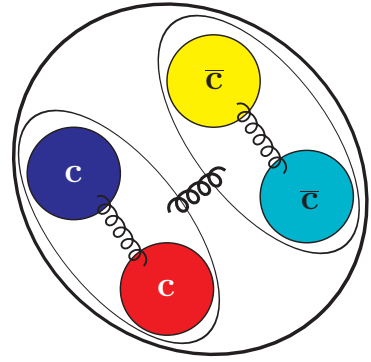}
\captionof{figure}{$T_{4c}$ in diquark-antidiquark picture.}
\end{minipage}%
\end{minipage}

\section{Formalism}

We adopt a quarkonium model \cite{model}, solving the Sch\"{o}dinger equation \cite{notebook} with the potential $V^{(0)}(r)$, which has a Coulomb term from one gluon exchange, linear confinement and a smeared spin-spin interaction:
\begin{equation}
\label{V0}
    V^{(0)}(r) = \kappa_{s}\frac{\alpha_{s}}{r} + b r -\frac{8\pi\kappa_s\alpha_s}{3m_c^2}\left(\frac{\sigma}{\sqrt{\pi}}\right)^3\mathrm{e}^{-\sigma^2r^2}
    \mathbf{S_1 \cdot S_2}
\end{equation}
where the color factor $\kappa_s$ is $-4/3$ for color singlet and $-2/3$ for vector diquark in antitriplet color state, while the QCD strong coupling constant $\alpha_s$, string tension $b$ and gaussian parameter $\sigma$, together with the charm quark mass $m_c$, are fitted to the $c\bar{c}$ spectrum (Fig. \ref{charmonium}).
The spin-orbit and tensor interactions are included as first order perturbation:
\begin{equation}
\label{V1}
  V^{(1)}(r) = \left( -\frac{3\kappa_s\alpha_s}{2m^2}\frac{1}{r^3}
    - \frac{b}{2m^2}\frac{1}{r} \right) \mathbf{L \cdot S}
      -\frac{\kappa_s\alpha_s}{4m^2} \frac{1}{r^3} \mathbf{S_{12}}
\end{equation}
\vspace{0.2cm}
where the operator
\begin{equation}
 \mathbf{S_{12}}
= 4[ 3(\mathbf{S_1 \cdot \hat{r}})(\mathbf{S_2 \cdot \hat{r}})  -\mathbf{S_1 \cdot S_2} ] \nonumber
\end{equation}
can be written in terms of spherical harmonics $Y_2^m$ and spin operators $S_+$, $S_-$, $S_z$ to be applied directly in the tetraquark wavefunction \cite{Debastiani}.
First we calculate the diquark mass using $V_{cc} = V_{c\bar{c}}/2$, where the attractive antitriplet color structure implies the diquark has total spin 1. Next, the diquark mass is used as input to obtain the tetraquark spectrum as a two-body system, allowing radial and orbital excitations between diquark and antidiquark.


\section{Results and Conclusions}

\begin{figure}[h!]
\centering
\includegraphics[width=0.95\textwidth]{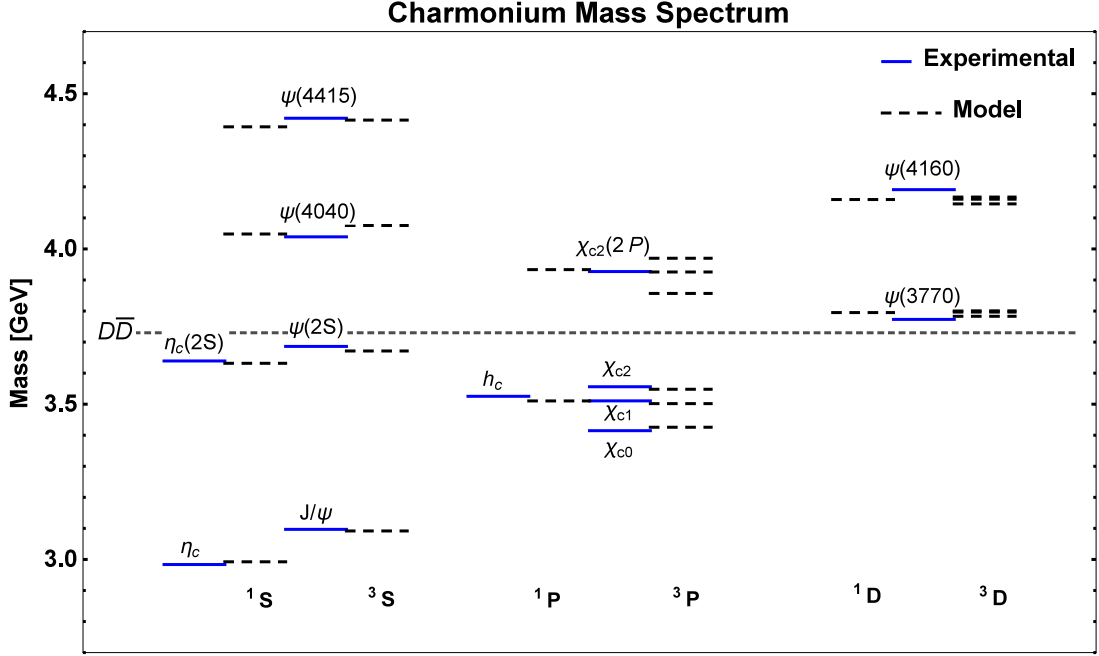}
\caption{\label{charmonium} A good reproduction of $c\bar{c}$ mass spectrum is obtained with $m_c = 1.4622$ GeV, $\alpha_s = 0.5202$, $b=0.1463$ GeV$^2$, $\sigma=1.0831$ GeV.}
\end{figure}

In Fig. \ref{charmonium} we compare the model results with recent experimental data of $c\bar{c}$ states.


We find that the non-relativistic approximation is reasonable,
with $\left\langle v^2/c^2 \right\rangle< 0.2$. The lowest $1S$ states of the $T_{4c}$ are found to be very compact, with $\langle r^2 \rangle^{1/2}<0.3$ fm, where the binding is mostly due to the Coulomb term from one gluon exchange, which also creates a big overlap of the wavefunction at the origin and a strong contribution of the spin-spin interaction.

On the other hand, the lowest diquark is found to have $\langle r^2 \rangle^{1/2}\sim 0.6$ fm, warning against the point-like approximation, which is very common in the literature, even though the two-body approximation should still be reasonable due to the color structure behind it.

In Fig. \ref{tetraquark} we show the results for the $T_{4c}$ mass spectrum in analogy to the $c\bar{c}$ one, considering one radial and one orbital excitation and the splitting structure from the spin-dependent terms, 
in a total of 20 states. 
We can see that the three lowest $1S$ states with $J^{PC}=0^{++}, \, 1^{+-},\, 2^{++}$ and masses 5969.4, 6020.9, 6115.4 MeV seem to be below\footnote{For self-consistence we use the $c\bar{c}$ meson masses calculated with the model. If experimental values were used, the $0^{++}$ state would actually be a few MeV above the $\eta_c$ pair threshold.} the two-meson thresholds $\eta_c\,\eta_c$, $\eta_c \,J/\psi$ and $J/\psi \, J/\psi$, respectively.
The inclusion of orbital momentum increases the possibilities in quantum numbers and decay channels. One interesting case is the $1^3P_1$ state with exotic quantum numbers $J^{PC}=1^{-+}$ and mass 6577.4 MeV, 
about 80 MeV above the corresponding $\eta_c(1S)\,\chi_{c1}(1P)$ threshold. Detailed results and discussions are presented in Ref. \cite{Debastiani}.

\noindent\begin{minipage}{\textwidth}
\begin{minipage}[c][10cm][c]{\dimexpr0.3\textwidth-0.5\Colsep\relax}
\scriptsize \label{tab:masses}
\captionof{table}{Results for $[cc]$ diquark and $T_{4c}$ tetraquark masses in MeV.}
\centering
\begin{tabular}{|c |r |l|}
\hline\hline
    $N^{2S+1}\ell_J$ & $M_{T4c}$ & $J^{PC}$ \\
\hline
   \multicolumn{3}{|c|}{\textbf{Diquark}}  \\
\hline
$1^3S_1$ & $3133.4$ & $1^+$ \\
\hline
   \multicolumn{3}{|c|}{\textbf{Tetraquark}}  \\
\hline
    $1^1S_0$ & $5969.4$ & $0^{++}$ \\
    $1^3S_1$ & $6020.9$ & $1^{+-}$ \\
    $1^5S_2$ & $6115.4$ & $2^{++}$ \\
\hline
    $1^1P_1$ & $6577.1$ & $1^{--}$ \\
\hline
    $1^3P_0$ & $6480.4$ & $0^{-+}$ \\
    $1^3P_1$ & $6577.4$ & $1^{-+}$ \\
    $1^3P_2$ & $6609.9$ & $2^{-+}$ \\
\hline
    $1^5P_1$ & $6495.4$ & $1^{--}$ \\
    $1^5P_2$ & $6600.2$ & $2^{--}$ \\
    $1^5P_3$ & $6641.2$ & $3^{--}$ \\
\hline
    $2^1S_0$ & $6663.3$ & $0^{++}$ \\
    $2^3S_1$ & $6674.5$ & $1^{+-}$ \\
    $2^5S_2$ & $6698.1$ & $2^{++}$ \\
\hline
    $2^1P_1$ & $6944.1$ & $1^{--}$ \\
\hline
    $2^3P_0$ & $6866.5$ & $0^{-+}$ \\
    $2^3P_1$ & $6943.9$ & $1^{-+}$ \\
    $2^3P_2$ & $6970.4$ & $2^{-+}$ \\
\hline
    $2^5P_1$ &  $6875.6$ & $1^{--}$ \\
    $2^5P_2$ & $6962.1$ & $2^{--}$ \\
    $2^5P_3$ & $6996.7$ & $3^{--}$ \\
\hline\hline
\end{tabular}
\end{minipage}\hfill
\begin{minipage}[c][10cm][c]{\dimexpr0.7\textwidth-0.5\Colsep\relax}
\centering
\includegraphics[width=\textwidth]{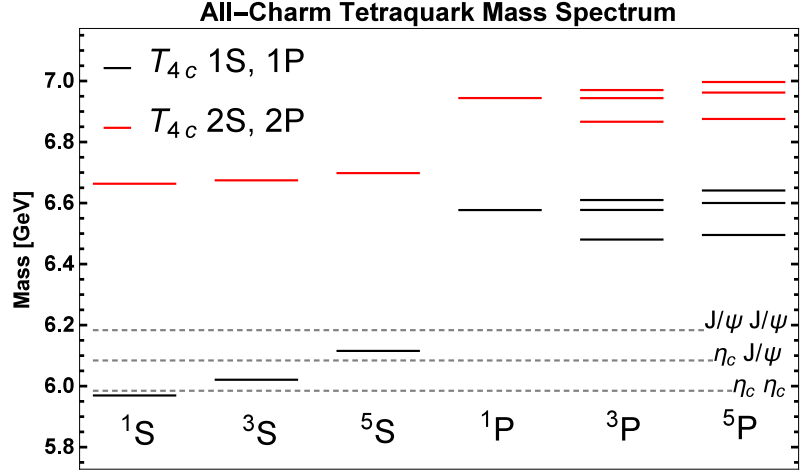}
\captionof{figure}{\label{tetraquark} Tetraquark spectrum including radial and orbital excitations relative to the diquark-antidiquark system and spin-dependent splitting. Ground state ($N^{2S+1}\ell_J=1^3S_1$) diquarks are used, with $m_{cc} = 3133.4$ MeV.}
\end{minipage}%
\end{minipage}%
\vspace{10pt}

\section{Acknowledgments}

The authors acknowledge the support received from the brazilian funding agencies
FAPESP (contract 12/50984-4), CNPq and CAPES.
We also thank Profs. W.~Lucha, F.~F.~Sch\"{o}berl for the \textsf{Mathematica} notebook; and M.~R.~Robilotta, E.~Oset, A. Valcarce, J.~Vijande for the fruitful discussions.
V. R. Debastiani also acknowledges the support from Generalitat Valenciana in the Program Santiago Grisolia (Exp. GRISOLIA/2015/005) and wishes to thank the organizers of the Hadron 2017 for the award this work received in the poster section and S. Neubert, I. Belyaev for the interest and invitation to present this work for LHCb Collaboration in the Hadron Spectroscopy Phenomenology Workshop.


\begin{thebibliography}{99}


\bibitem{Debastiani} {
   V.~R.~Debastiani and F.~S.~Navarra,
  ``A non-relativistic model for the $[cc][\bar{c}\bar{c}]$ tetraquark,''
  arXiv:1706.07553 [hep-ph].}

\bibitem{Iwasaki}          Y.~Iwasaki, {Prog.\ Theor.\ Phys.\ {\bf 54}, 492 (1975)};
  {UTHEP-2 (1975)};
  {Phys.\ Rev.\ Lett.\ {\bf 36}, 1266 (1976)};
  {Phys.\ Rev.\ D {\bf 16}, 220 (1977).}


\bibitem{Chao}             K.~T.~Chao, {Z.\ Phys.\ C {\bf 7}, 317 (1981).}

\bibitem{model}            T.~Barnes, S.~Godfrey, E.~S.~Swanson,
                           {Phys.\ Rev.\ D {\bf 72}, 054026 (2005).}

\bibitem{notebook}            W.~Lucha, F.~F.~Sch\"{o}berl,
                           {Int.\ J.\ Mod.\ Phys.\ C {\bf 10}, 607 (1999).}


\end{thebibliography}
\end{document}